# Statistical Contextual Explanation of Quantum Paradoxes

**Marian Kupczynski**

Département de l'Informatique et d'Ingénierie, UQO, Case postale 1250, succursale Hull, Gatineau. QC, Canada J8X 3X7

**\* Correspondence:**
marian.kupczynski@uqo.ca

**Abstract**

We celebrate this year hundred years of quantum mechanics. Incorrect interpretations of QM and incorrect mental models of invisible details of quantum phenomena lead to paradoxes. In this review article we advocate the statistical contextual interpretation (SCI) of quantum mechanics. State vectors (wave functions) and various operators are purely mathematical entities allowing making quantitative probabilistic predictions. State vector describes an ensemble of identically prepared physical systems and a specific operator represents a class of equivalent measurements of a physical observable. A collapse of wave function is not a mysterious and instantaneous physical process. A collapsed quantum state describes a new ensemble of physical systems prepared in a particular way. A value of a physical observable, such as a spin projection, associated with a pure quantum ensemble is a characteristic of this ensemble created by its interaction with measuring instruments. Probabilities are objective properties of random experiments in which empirical frequencies stabilize. SCI rejects claim that quantum mechanics provides a complete description of individual physical systems but it remains agnostic on whether a more detailed subquantum description can be found or is necessary. In conformity with Bohr contextuality, SCI rejects Bell-local and Bell-causal hidden variable models. Nevertheless, by incorporating into probabilistic model contextual hidden variables measuring instruments, long distance quantum correlations studied in Bell Tests can be explained without evoking quantum nonlocality or retro-causality. SCI does not claim to provide a complete description of quantum phenomena. In fact, we even don't know whether quantum probabilities provide a complete description of existing experimental data. Time series of experimental data may contain much more information than it is obtained using empirical frequencies and histograms. Therefore, predictable completeness of quantum mechanics has be tested and not taken for granted.

## 1. Introduction

In 1925, Werner Heisenberg, Max Born and Pascual Jordan developed Matrix Mechanics [1-3], the first consistent formulation of quantum mechanics. To commemorate this achievement, 2025 has been declared the International Year of Quantum Science and Technology (IYQ) by the United Nations.

In spite of the incredible advances made in quantum science and technology over the past century there is still no consensus regarding its interpretation and limitations [3-6]. Incorrect interpretations of QM and incorrect mental models of invisible details of quantum phenomena lead to paradoxes and speculations about quantum weirdness and quantum magic. Most of these paradoxes are due to the *individual interpretation* according to which an instantaneous collapse of wave function describing individual physical system(s) is triggered by a single measurement performed on one of these systems.

In this article, we review and advocate a statistical contextual interpretation (SCI) [7-23] which is free of paradoxes. According to this interpretation a quantum state is not an attribute of an individual physical system which may be changed instantaneously. The so called collapse of the wave function it is not a mysterious physical process. Quantum state/ wave function is a mathematical entity representing an equivalence class of subsequent preparations of the physical systems. Quantum states together with specific operators representing physical observables are used to make probabilistic predictions for a statistical scatter of measured values of these observables. SCI [11-14] is similar but not identical to Ballentine's statistical [9, 10] and Khrennikov's Växjö interpretation [15-18].

A probability can have a different meaning [15, 24]. In SCI, it is an objective property of a random experiment in which empirical frequencies stabilize, thus a probabilistic description of quantum phenomena can hardly be considered as a complete description of individual physical systems [8-11]. This is why, Einstein believed that QM is an emergent theory and that a more detailed description of quantum phenomena should be found [7-9]. Bohr insisted that quantum probabilities were irreducible and that QM provided a complete description of quantum phenomena and experiments [25-28].

In 1927, Heisenberg [29] demonstrated the uncertainty principle according to which one may not measure simultaneously, with arbitrary accuracy, a linear momentum $p$ and a position x of a sub-atomic particle : $\Delta x \Delta p \leq h$ where $h$ is a Planck constant. The principle was generalised by Robertson [30] and its precise statistical meaning was given by Kennard [31]. We have two experiments performed on two identically prepared beams/ensembles of "particles". In one experiment we measure their linear momenta and in another their positions. A statistical scatter of experimental data is described by respective standard deviations and: $\sigma_x \sigma_p \leq \hbar/2$ where $\hbar = h/2\pi$. In this interpretation we are talking only about a statistical scatter of measurement outcomes and not about positions and linear momenta of "particles" if no measurements are done. According to Copenhagen interpretation (CI), all the speculations about sharp unmeasured values of linear momenta and positions of sub-atomic particles are meaningless and QM does not imply that <u>an electron can be here and a meter away at the same time</u>, what is incorrectly claimed by several authors [6].

In 1935, Einstein, Podolsky and Rosen [32], proposed a thought experiment, called the EPR paradox, intended to demonstrate the incompleteness of quantum mechanics. They considered two entangled particles which interacted in the past moving away from each other in distant locations. According to Copenhagen interpretation (CI), measuring the position or momentum of one particle would instantly give the information about the position or momentum of its distant partner without disturbing it in any way. Thus physical properties of objects exist independently of measurement contrary to CI. Bohr explained that EPR inference requires different incompatible, but complementary, experiments and it could not provide more information about an individual physical system, than it was allowed by QM [33].

EPR paradox was rephrased by Bohm [34] in terms of measurements of particle's spin. If you measure the spin of one particle, you instantly know the spin of the other. According to QM outcomes are produced in an irreducibly random way but in an ideal EPR-B experiment, they are perfectly correlated or anti-correlated in specific experimental settings. This is called the EPR-B paradox: since: *a pair of fair dice cannot produce always correlated outcomes* [35-38].

Bell [39, 40] abandoned irreducible randomness and proposed Local Realistic Hidden Variable Model ( LRHVM) in which outcomes were predetermined at a source. Clauser and Horne [41] abandoned predetermination and proposed Stochastic Hidden Variable Model ( SHVM). LRHVM describes entangled pairs/qubits as pairs of socks and SHVM as a pair of dice. In these models, correlations between distant outcomes coded ±1 have to obey Bell-CHSH inequalities [42]. Later, hidden variables were assumed to represent all common causes of events in distant laboratories and Local Hidden Variable Model (LHVM) [36, 40, 43, 44] could be rejected in several Bell Tests [45-52].

Since Bell-CHSH inequalities are violated by some quantum predictions and by experimental data the majority of physical community believes that no other locally causal explanation of quantum correlation is possible. Therefore, nature does exhibit non-locality and entangled particles can influence each other instantaneously across huge distances. This is a source of extraordinary metaphysical speculations about *experimenters' freedom of choice, retro-causality* and *quantum nonlocality*.

It was explained by several authors, that such speculations were unfounded
[5, 6, 11-23, 37, 53-101.

- In spin polarisation correlation experiments (SPCE) and other Bell Tests for different pairs of settings we have 4 incompatible random experiments. LRHVM are using a unique probability space and a joint probability distribution to describe these experiments what is only possible in rare circumstances and what is clearly incompatible with experimental protocols in Bell Tests [12-15, 21-23, 37, 53-58, 61, 63, 65-69, 72-76, 85, 93, 94].

- Bell and CHSH inequalities are trivial algebraic properties of experimental spreadsheets [38, 65, 81, 98-100] containing quadruplets of ±1, which are in fact samples drawn from a statistical population described by some joint probability distribution of 4 compatible random variables. The outcomes of Bell Tests are displayed using 4 spreadsheets containing each only couples ±1. The violation of Bell-CHSH inequalities provides in fact only the evidence that the data in these 4 spreadsheets cannot be reshuffled to form quadruples [99,100].

- In QM, interactions of instruments with physical systems during the measurement process may not be neglected and outcomes are not passively registered pre-existing values of the physical observables. Therefore, Bell-causal hidden variable model suffer from theoretical "*contextuality loophole*" [23, 37, 38, 82-85, 88-90], because they fail to include correctly setting dependent contextual variables describing measuring instruments at the moment of a measurement.

A detailed discussion of EPR-type paradoxes and Bell Tests in the spirit of SCI may be found in [11-13] and in a dedicated section of this article. As we conclude in [84, 85], Bell Tests allow only rejecting probabilistic couplings provided by Bell-local and Bell-causal hidden variable models. If contextual variables, describing varying experimental contexts, are correctly incorporated into a probabilistic model, then Bell–CHSH inequalities cannot be proven and *nonlocal quantum correlations* may be explained in an intuitive way.

The paper is organised as follows. In section 2 we recall different definitions of probability and Bertrand`s paradox suggesting that in physics probabilities are objective properties of random experiments in which empirical frequencies stabilize. In section 3 we compare classical and quantum observables and filters. In section 4 we recall EPR-B paradoxes and we explain them using SCI. In section 5 we derive quantum predictions for an ideal EPR-B experiment. In section 6 we explain that Bell-CHSH inequalities are trivial arithmetic properties of Nx4 spreadsheets containing ±1 entries and that they can be rigorously derived only for random experiments described by 4 binary jointly distributed random variables. In section 7 we discuss hidden variable models proposed to explain EPR-B experiments. Section 8 is about loophole free Bell Tests, their interpretation and implications. In section 9 we present a contextual hidden variable model, which allows explaining long range correlations observed in Bell Tests. In section 10 we advocate more detailed analysis of existing time-series of data in order to elucidate the problem of completeness of quantum mechanics. Section 11 contains some additional conclusions.

1. **Probability and Bertrand paradox**

Probability and randomness are subtle notions which have been debated by mathematicians and philosophers for centuries. There exist several definitions of probability [15, 24].

Classical probability is the ratio of the number of favorable outcomes to the total number of possible outcomes. For example the probability of drawing a black king from a deck of 52 cards is 2/52=1/26. Geometric probability is the probability that a point chosen at random within a certain geometric figure will satisfy a given condition and is calculated as the ratio of the area (or length, volume, etc.) of the favorable region to the area of the entire region. For example the probability of hitting a specific region on the dartboard can be calculated by dividing the area of that region by the total area of the dartboard.

Frequentist probability is the relative frequency of occurrence of an experiment's *outcome* "in the long run" of outcomes (theoretically if experiment would be repeated infinite number of times). It is an objective property of a random experiment. Another objective probability is *propensity* which is defined as the tendency of some experiment to yield a certain outcome, even if it is performed only once. A subjective probability is based on a personal judgment of an agent and quantifies her degree of belief how likely an event is to occur.

The limitations of the classical and geometric probabilities became evident due to the Bertrand's Paradox. It demonstrates how different methods of defining "randomness" can lead to different probabilities for the same event. In 1889, Bertrand posed the following problem: Consider an equilateral triangle inscribed in a circle. What is the probability that a randomly chosen chord of the circle is longer than a side of the triangle? He provided three different methods to choose a random chord, each yielding a different probability [102, 103].

Bertrand paradox can be rephrased in a more intuitive way [21]. If we consider two concentric circles on a plane with radii R and R/2 respectively, we can ask a question what is the probability P that a chord of the bigger circle chosen at random cuts the smaller one in at least one point? The various answers seem to be equally reasonable: if we divide the ensemble of all chords into sub-ensembles of parallel chords, we find P= 1/2. If we consider sub-ensembles of chords having the same beginning, we find P=1/3. Finally if we choose midpoints of chords lying in the small circle, we find P=1/4.

The solution of the paradox is simple. Different probabilistic models leading to different answers correspond to random experiments performed using different specific experimental protocols. It proves contextual character of probabilities and their intimate relation to specific random experiments [23]. Therefore, the probability of obtaining 'head' in a coin flipping experiment using a specific coin and a specific flipping device is neither a property of the coin nor a property of the flipping device. It is only a property of the whole experiment: "flipping this particular coin with that particular flipping device". This is why in physics probabilities are objective properties of phenomena and random experiments in which empirical frequencies stabilize.

## 2. Classical versus quantum: properties, filters and observables.

In classical physics, measurement outcomes may contain experimental errors but measurements are assumed to be non-invasive, what means that they do not change properties they measure. Therefore, macroscopic physical systems are described by properties $p_i$ (i=1,..,n) quantified by the values of classical compatible observables which can be measured in any order.

If we have a mixed statistical ensemble (a beam) B of macroscopic systems we can choose systems having particular properties using classical filters. *A classical filter $F_i$ or a macro selector is a device which passes through only systems having a property $p_i$. Classical filters operate according to Boolean Yes- or -No logic. If we have n different properties we have n filters corresponding to them. A lattice of classical filters have simple properties: $F_i F_i = F_i$, $F_i F_j = F_j F_i$ There exists also a maximal filter $F = F_1 F_2 ...F_n$ which transforms a mixed statistical ensemble into a pure statistical ensemble in which all the systems have exactly the same properties.* [23]. Mixed statistical ensemble of physical systems can be described by a joint probability distribution of random variables associated with measured physical observables.

In quantum experiments, the information obtained about invisible physical systems is indirect and obtained from their interactions with macroscopic measuring instruments. As Bohr correctly insisted, the atomic phenomena are characterized *"the impossibility of any sharp separation between the behaviour of atomic objects and the interaction with the measuring instruments which serve to define the conditions under which the phenomena appear"* (Bohr ([26], v. 2, pp. 40–41). Quantum observables have the following properties [104]: *Bohr-contextuality: The output of any quantum observable is indivisibly composed of the contributions of the system and the measurement apparatus.* [84]

The formalism of QM was inspired by optical experiments with polarized light. Linearly polarized light passes without noticeable attenuation by a subsequent identical polarizer. The intensity of linearly polarised light after a passage through another polarizer is reduced according to Malus law: $I = I_0 \cos^2 \theta$, where $I_0$ is the initial intensity and θ is the angle between the light's initial polarization direction and the axis of the polarizer.

Discrete atomic spectral lines and photoelectric effect proved that exchanges of energy between electromagnetic field and matter are quantized and "carriers" of quantized exchanged energy are called *photons*. Therefore, linearly polarized monochromatic light is usually represented as a beam of linearly polarized *photons* carrying energy hν. This mental picture is misleading because we cannot see photons and they are not point-like objects. When a sophisticated photon detector, after several steps of signal enhancement, produces a click we conclude: *a photon was detected*. The intensity of light is now measured by counting clicks on detectors and we say that each linearly polarized photon has a probability ( propensity) $p = \cos^2 \theta$, to pass through a polarizer, if θ is the angle between the direction of photon's initial polarization and the axis of the polarizer.

After passing by quantum filter , linear polarisation of light becomes a contextual property of photons. *A quantum filter $F_i$ , it is a device which creates a contextual property "i": "passing by $F_i$". A physical system having a property " i " have a probability ( propensity) $p_{ij}$ to pass by another filter $F_j$ acquiring after the passage a new property " j " . Quantum filters are idempotent $F_i F_i = F_i$ but in general they do not commute $F_i F_j \neq F_j F_i$ and the lattice of quantum filters is isomorphic to the lattice of projectors on subspaces of a Hilbert space. Quantum filters are not selectors of pre-existing attributes of physical systems but are creators of contextual properties defined above.* [23]

Incompatible filters, such as polarizers with non-parallel axes, create incompatible contextual properties which cannot be measured simultaneously and if measured in a sequence a previous contextual property is destroyed in a new measurement. As explained in the preceding section the probabilities are objective properties of phenomena and random experiments thus talking about propensity as the property of individual physical systems (here invisible photons) is in fact unfounded. This is why in SCI quantum state vectors are not considered to be properties of the individual physical systems. Treating a wave function as an attribute of the individual physical system leads to EPR paradox which we discuss in the next section.

4. **EPR Paradox and statistical contextual interpretation**

We resume below the discussion of EPR paradox we gave in [13]. Before, the publication of EPR paper it was believed that:

   A1: Any pure state of a physical system is described by a specific <u>unique</u> wave function $\Psi$.

   A2: Any measurement causes a physical system to jump instantaneously into one of eigenstates of the dynamical variable that is being measured. This eigenstate becomes a new wave function describing a state of the system.

   A3: A wave function $\Psi$ provides a <u>complete</u> description of a pure state of an <u>individual</u> physical system.

EPR consider two particular individual systems I+II in a pure quantum state, which interacted in the past, separated and evolved freely afterwards [32]. Using A2 they conclude that:

- A single measurement performed on one of the systems, for example on the system I, gives instantaneous knowledge of the wave function of the system II moving freely far away.

- By choosing two different incompatible observables to be measured on the system I it is possible to assign two different wave functions to the system II (the same physical reality: the second system after the interaction with the first).

Since a measurement performed in a distant location on the system I does not disturb in any way the system II, thus according to A1 and A3 the system II should be described by a unique wave function not by two different wave functions. Moreover these wave functions are eigenstates of two non-commuting operators representing incompatible physical observables what allows to deduce indirectly the values of these incompatible physical observables for the same system II without disturbing it in any way what contradicts Heisenberg uncertainty relations and CI.

EPR discussed particles' positions and momenta, Bohm discussed an experiment in which a source produces pairs of particles prepared in a spin singlet state [34]. One member of a pair (photon or electron) is sent to Alice and another to Bob in distant laboratories. According to A1, each pair of photons is described by a state vector:

$$\Psi = (|+\rangle_P |-\rangle_P - |-\rangle_P |+\rangle_P)/\sqrt{2}. \tag{1}$$

where $|+\rangle_P$ and $|-\rangle_P$ are state vectors corresponding to photon states in which their spin is "up" or "down" in the direction **P** respectively. If we measure a spin projection of a photon I on the direction **P** we have an equal probability to obtain a result "1" or "–1". If we obtain "1" a reduced state vector of the photon II is $|-\rangle_P$, if we obtain "-1" a reduced state vector of the photon II is $|+\rangle_P$. By choosing a direction **P**, for the measurement to be performed on the photon I, when " photons are in flight and far apart" we can assign different incompatible reduced state vectors to the same photon II. In other words: we can predict with certainty, and without in any way disturbing the second photon, that **P**-component of spin of the photon II must have the opposite value to the value of the measured **P**- component of the spin of the photon I. [10] Therefore for any direction **P** the **P**-component of the spin of the photon II has unknown but predetermined value what contradicts QM and is called the EPR-B experiment of paradox.

Bohr [33] promptly replied to EPR paper and explained that two different wave functions could be assigned to the system II only in two different incompatible experiments in which both systems were exposed to different influences before the measurement on the system I was performed . In order to be able to make predictions concerning individual physical systems in EPR scenario one has to have much more detailed knowledge how a particular pair was prepared in each of these incompatible experiments [11].

In 1936, Einstein advocated purely statistical interpretation of QM and explained that EPR paradox disappears because: "*Ψ function does not, in any sense, describe the state of one single physical system and reduced wave functions describe different sub-ensembles of systems*" [8]. This statistical interpretation has been generalized and promoted with success by Ballentine [9, 10] : "*the habit of considering an individual particle to have its own wave function is hard to break ...though it has been demonstrated strictly incorrect*" .

According to the statistical contextual interpretation of QM (SCI) [10, 11-13, 17, 19]:

1. A state vector Ψ is not an attribute of a single electron, photon, trapped ion, quantum dot etc. A state vector Ψ or a density matrix ρ describe only an ensemble of identical state preparations of some physical systems

2. A wave function reduction is neither instantaneous nor non-local. In EPR experiment a state vector describing the system II obtained by reduction of an entangled state of two physical system I+II describes only an sub-ensemble of systems II being partners of those systems I for which a measurement of some observable gave the same specific outcome. Different sub-ensembles are described by different reduced state vectors.

3. A value of a physical observable, such as a spin projection, is not a predetermined attribute of a system but it is a property of a pure ensemble of identically prepared physical systems created in the interaction with a measuring instrument [23, 77].

The solution of EPR-B paradox given by SCI is simple: the wave function reduction is not instantaneous and a reduced one particle state $|+\rangle_P$ "describes" only an ensemble of partners of the particles I which were detected to have "spin down" in the direction **P.** For different directions **P,** we perform specific experiments and we obtain different sub- ensemble of particles II. Strong correlations between distant outcomes in EPR experiments are due to various conservation laws. More detailed discussion of EPR and EPR-B paradoxes may be found for example in [11-13**].**

## 5. Kolmogorov and quantum probabilistic models .

Outcomes of any random experiment are described by a specific probability space Ω, σ- algebra of its all sub-ensembles F and a probabilistic measure μ. A sub-ensemble $E \in F$ is an event corresponding to a subset of possible outcomes of a random experiment. A probability of observing this event is given by $0 \leq \mu(E) \leq 1$. In statistics instead of Ω we use a sample space S which contains <u>only possible outcomes</u> of a studied random experiment.

Every random experiment is defined by its experimental context C. [14-18, 23, 76]. If its outcomes are discrete it may be described by a random variable A and a probability distribution:

$$P(a|C) = P(A = a|C). \tag{2}$$

and its expectation value

$$E(A|C) = \sum_a aP(a|C) . \tag{3}$$

In quantum experiments a context of an experiment is determined by a preparation of an ensemble of physical systems represented a density operator ρ (or a state vector ψ) and by a Hermitian operator $\hat{A}$ representing experimental set-up used to measure a physical observable A. Instead of (2) and (3) we have:

$$P(a|\psi,\hat{A}) = |\langle a|\psi\rangle|^2, \qquad (4)$$

where $|a\rangle$ is a corresponding eigenvector of the operator $\hat{A}$ and

$$E(A|\psi,\hat{A}) = \langle \psi|\hat{A}|\psi\rangle. \qquad (5)$$

If a density matrix $\rho$ is used to described a pure or mixed prepared ensemble then:

$$E(A|\rho,\hat{A}) = Tr(\rho\hat{A}) \qquad (6)$$

In an idealized EPR-B experiment (1) impossible to implement, a source is sending two correlated signals which arrive to distant laboratories, pass by polarization analyzers and after arriving to detectors produce coincident clicks on them. The experimental situation is much more complicated since clicks are not registered at the same time and one has to decide which clicks are correlated by introducing specific time windows and deciding how to use them in order to define coincident clicks [37, 82].

An idealized EPR-B experiment is described by a following probabilistic model [38, 83, 84, 105]. Randomly chosen polarisation measurement settings are (x, y), prepared ensemble E is described by $\rho = |\psi\rangle\langle\psi|$, $\hat{A}_x = \vec{\sigma}\cdot\vec{n}_x$ and $\hat{B}_y = \vec{\sigma}\cdot\vec{n}_y$ represent spin projections on the corresponding unit vectors and:

$$E(A_x B_y) = Tr(\rho\hat{A}_x \otimes \hat{B}_y) = \langle\psi|\hat{A}_x \otimes \hat{B}_y|\psi\rangle = \sum_{\alpha\beta}\alpha\beta p_{xy}(\alpha,\beta) = -\vec{n}_x\cdot\vec{n}_y = -\cos(\theta_{xy}) \quad (7)$$

where $\hat{A}_x \otimes \hat{B}_y |\alpha\beta\rangle_{xy} = \alpha\beta|\alpha\beta\rangle_{xy}$, $p_{xy}(\alpha,\beta) = |\langle\psi|\alpha\beta\rangle_{xy}|^2$ and $\alpha=\pm 1$ and $\beta=\pm 1$ [84,105].

The model is contextual because a triplet $\{\rho,\hat{A}_x,\hat{B}_y\}$ changes, if a preparation or experimental settings change. For each choice of settings (x, y), QM provides a specific Kolmogorov model.

Since $E(A_x B_y) = -1$ for $\theta_{xy} = (\theta_x - \theta_y) = 0$, it has been incorrectly claimed that QM predicts strict anti-correlations of two space-like events produced in an irreducible random way. Since two space-like events produced in an irreducible random way cannot be correlated ($E(A_x B_y) = 0$), in absence of spooky influences, thus irreducible randomness was abandoned and several hidden variable models were proposed trying to explain strong long range correlations in EPR-type experiments predicted by QM . We will discuss these models later, but first in the next section we define and explain Bell-CHSH inequalities which play an important role in the discussions about the completeness of quantum mechanics. Before doing it, we point out that for Bell Tests there are no strict correlations between distant outcomes, predicted by quantum mechanics, Directions can only be defined by some small intervals $I_x$ and $I_y$ containing angles close to $\theta_x$ and $\theta_y$ respectively. Therefore, the correct quantum prediction for expectation values is [13,77]:

$$E(A_x B_y) = -\iint_{I_x I_y} \cos(\theta_1 - \theta_2) \, d\rho_x(\theta_1) \, d\rho_y(\theta_2). \tag{8}$$

### 6. Experimental spreadsheets and Bell –CHSH inequalities.

Let us consider a random experiment described by 4 jointly distributed binary random variables (A, A', B, B') taking the values ±1. In each trial of this experiment 4 outcomes (a. a, b, b') are obtained and displayed in a Nx4 experimental spreadsheet [38]. Since b=b' or b= -b'thus

$$|s| = |ab - ab' + a'b + a'b'| = |a(b-b')| + |a'(b+b')| \leq 2. \tag{9}$$

From (9) we obtain immediately CHSH inequality:

$$|S| \leq \sum_{a,a',b,b'} |ab - ab' + a'b + a'b'| \, p(a,a',b,b') \leq |E(AB) - E(AB')| + |E(A'B) + E(A'B')| \leq 2 \tag{10}$$

where p(a, a, b, b') is a joint probability distribution of (A, A', B, B' ) and $E(AB) = \sum_{a,b} ab \, p(a,b)$ is a pairwise expectation of A and B obtained using a marginal probability distribution $p(a,b) = \sum_{b,b'} p(a,a',b,b')$ [38].

If all pair-wise expectation values in (10) are estimated using the same Nx4 experimental spreadsheet then the inequality (10) is strictly obeyed by all finite samples. The inequalities (10) are in fact necessary and sufficient conditions for the existence of a joint probability distribution of only pairwise measurable ±1-valued random variables [63]. The inequalities (10) are also valid if $|A|\leq 1$, $|A'|\leq 1$, $|B|\leq 1$ and $|B'|\leq 1$. It is now well known that cyclic combinations of pairwise marginal expectations of jointly distributed binary random variables must obey non-contextuality inequalities (NCI) [106]. Bell-CHSH inequalities are a special case of NCI .

If we have, four Nx4 spreadsheets containing outcomes from 4 runs of the same random experiment , discussed above, but we use each of these spreadsheets to estimate only one pairwise expectation E (A, B), E(A,B'), E (A'. B) and E(A'. B') respectively, then 50% of time these estimates violate the inequality (10) [13, 83, 107]. Only if, N is increasing to infinity the probability of the violation the inequality (10) tends to 0. Therefore the violation of CHSH-inequality by experimental data in EPR-type experiments allows only to evaluate a plausibility of particular probabilistic models [84]. In next section we are going to discuss such models.

### 7. Local Realistic Models for EPR-Bohm Experiment

i) Local realistic hidden variable model (LRHVM)

In attempt to explain correlations in an ideal EPR-B experiment Bell [23, 39, 40, 84] proposed a probabilistic model in which outcomes registered in distant laboratories are predetermined at a source:

$$E(A_x B_y) = \sum_{\lambda \in \Lambda} A_x(\lambda) B_y(\lambda) P(\lambda) \qquad (11)$$

where $A_x(\lambda) = \pm 1$ and $B_y(\lambda) = \pm 1$. In LRHVM, we have 4 jointly distributed random variables $(A_x(L), B_y(L), A_{x'}(L), B_{y'}(L))$ being functions of the same random variable L. The random variable L is describes a classical random experiment in which $\lambda$ is sampled with replacement from a probability space $\Lambda$. For each value of $\lambda$, all outcomes are be calculated. LRHVM describes *entangled pairs* as pairs of socks, which can have different sizes and colours; e.g. Harry draws a pair of socks, sends one sock to Alice and another to Bob, who in function of (x, y) record corresponding properties colour or size.

Since $(A_x(L), B_y(L), A_{x'}(L), B_{y'}(L))$ are jointly distributed, thus they obey CHSH inequality:

$$|S| = |E(A_x B_y) + E(A_x B_{y'}) + E(A_{x'} B_y) - E(A_{x'} B_{y'})| \leq 2. \qquad (12)$$

Bell knew very well that in EPR-B experiment $(A_x, B_y, A_{x'}, B_{y'})$ are not jointly measurable and that their joint probability distribution does not exist. He did not notice that to prove his inequalities he was tacitly using the existence of a joint probability distribution of $(A_x(L), B_y(L), A_{x'}(L), B_{y'}(L))$. As we explained in the preceding section (10) and (12) can be only be rigorously proven for a random experiment outputting in each trial four ±1 outcomes.

ii). Stochastic hidden variable model (SHVM)

Clauser and Horne [41, 84] proposed, a stochastic hidden variable model (SHVM), in which $\lambda$ does not determine outcomes in a given trial but only their probability. Using the notation of Big Bell Test collaboration [49]:

$$P(a,b \mid x, y) = \sum_{\lambda} P(a \mid x, \lambda) P(b \mid y, \lambda) P(\lambda) \qquad (13)$$

where P(-|-) denotes a conditional probability. The equation (13), for a fixed setting (x, y) describes a family of independent random experiments labelled by $\lambda$ and:

$$E(A_x B_y) = \sum_{\lambda} E(A \mid x, \lambda) E(B \mid y, \lambda) P(\lambda) \qquad (14)$$

Pair-wise expectations defined by (13) are also constrained by CHSH inequalities (12). In SHVM entangled photon pairs were described as pairs of dice and correlations, which might had been created in this way, were quite limited.

iii) Local causal hidden variable model (LHVM)

LHVM is a generalisation of preceding two models . In this model, $\lambda$'s represent all possible common causes of space like events happening in distant laboratories and: "*they may include the usual quantum state; they may also include all the information about the past of both Alice and Bob. Actually, the $\lambda$'s may even include the state of the entire universe*" [49, 84], except that inputs (x, y) cannot depend on them.

$$P(a, b, x, y) = \sum_{\lambda} P(a \mid x, \lambda) P(b \mid y, \lambda) P(x, y \mid \lambda) P(\lambda) \qquad (15)$$

and

$$P(x, y \mid \lambda) = P(x, y). \qquad (16)$$

The condition (16) is called *measurement independence*, experimenters' *freedom- of- choice* (FoC) or *no conspiracy* [49, 108-112]. Since correlation does not mean causation this terminology is based on the incorrect causal interpretation of conditional probabilities [14, 82-85,111,112]. In a probabilistic model $P(x, y \mid \lambda) \neq P(x, y)$ does not imply that FoC is constrained by causal influences.

If $\lambda$'s represent ontic properties of entangled pairs or common causes thus they cannot not depend in any sense on chosen settings:

$$P(\lambda, x, y) = P(\lambda) P(x, y) \Rightarrow P(\lambda \mid x, y) = P(\lambda) . \qquad (17)$$

However, hidden variables can describe also measuring instruments thus they can depend on the chosen settings [11-13]. As Theo Nieuwenhuizen explained the model (13) suffered from theoretical *contextuality loophole* because setting hidden variables describing measuring instruments had not been included [88-90].

We have no doubts that experimenters can freely choose binary random labels of their setting (x, y) and this is what they do [45-52]. However, this random choice of labels (x, y) is followed by a choice of corresponding specific instruments and setting dependent measuring procedures. Since measuring instruments play an active way in quantum experiments thus it is reasonable to assume that outcomes depend not only on setting independent hidden variables describing prepared physical systems but also on setting dependent hidden variables describing local instruments and measuring procedures and *statistical independence* (14) is violated:

$$P(\lambda \mid x, y) \neq P(\lambda) \qquad (18)$$

Bell was the first to notice that if hidden variables depended on settings, then Bell-CHSH inequalities could not be derived. However, since (18) implied the violation of (16) it was incorrectly believed that it would mean the violation of FoC [14, 49, 109, 83-85].

Bell clearly demonstrated that LRHVM is inconsistent with QM, because there exist 4 particular experimental settings for which using (7) one obtains: $|S| \leq 2\sqrt{2}$, which significantly violates (12). Various Bell Tests [45-52] were performed in order to check the plausibility of local hidden variable models. Before explaining a contextual hidden variable model in which the hidden variables depend on settings we discuss in the next section recent Bell Tests and their true implications.

### 8. Bell Tests and what have they proven.

Bell Tests are inspired by an ideal EPR experiment. Entangled pairs are created at a source and sent to distant locations or they are created directly in distant laboratories using specific synchronized preparations/treatments such as entanglement swapping or entanglement transfer protocols [84]. In spite of differences, experimental protocols are subdivided into 3 steps:

1) <u>Preparation of an ensemble E</u> of pairs of entangled physical systems.
2) <u>Random local choice of labels/ inputs (x, y)</u> using random number generators (RNG), and signals coming from the distant stars[48, 49] or /and human choices [49-52] . In this article we use 4 pairs of labels/inputs: (x, y), (x, y'), (x', y) and (x', y'), which denote 4 incompatible experimental settings/contexts.
3) <u>Implementation of correlated and synchronized measurements</u> in distant locations and readout of binary outcomes (a, b) (called outputs), which are the coded information corresponding to clicks on different distant detectors etc.

In Bell Test, in each trial, we perform 2 local independent random experiments to choose inputs. For each input corresponds a specific pair of correlated distant experiments. Outcomes of these experiments may be are described by 4 pairs of binary random variables: $(A_{xy}, B_{xy})$, $(A_{xy'}, B_{xy'})$, $(A_{x'y}, B_{x'y})$ and $(A_{x'y'}, B_{x'y'})$ [84]. Our notation is inspired by Contextuality by Default approach (CbD) [82,113-115] in which random variables measuring the same content in a different context are a priori stochastically unrelated e.g. $A_{xy}$ and $A_{x'y}$. It is evident that in Bell Tests a joint probability distribution of these 8 random variables does not exist and Bell- CHSH inequalities cannot be derived without additional assumptions [76].

A pair of random empirical variables ($A_{xy}$, $B_{xy}$) describes a scatter of outputs in the experiment using the settings (x, y). We have 4 random experiments described by specific empirical probability distributions. Using these empirical probability distributions we may test plausibility of quantum and local hidden variable models proposed to explain a statistical scatter of outcomes in an ideal EPR-B experiment. If random variables in probabilistic models are denoted ($A'_{xy}$, $B'_{xy}$), in order to not confound them with empirical random variables ($A_{xy}$, $B_{xy}$), then we say that a probabilistic model provides a *probabilistic coupling* if:

$$E(A_{xy})=E(A'_{xy}) \; , \; E(B_{xy})= E(B'_{xy}) \; , \; E(A_{xy}B_{xy})=E(A'_{xy}B'_{xy}) \qquad (19)$$

Therefore, in Bell Tests we are testing a plausibility of different probabilistic couplings, in particular for LRHVM :

$$E(A'_{xy})= E(A'_{xy'}) =E(A_x), \; , \; E(B'_{xy})= E(B'_{x'y})= E(B_y) \; , \; E(A'_{xy}B'_{xy})= E(A_xB_y) \; , \qquad (20)$$

where ($A_x$, $B_y$, $A_{x'}$, $B_{y'}$) are jointly distributed (11). More detailed discussion may be found in [84].

There is still a lot of confusion in the literature and on the social media concerning metaphysical implications of results of Bell Tests [49-52]. Using LHVM (15-17), one derives inequalities which have to be satisfied by specific combinations of probabilities of events to be observed in the experiments performed using different experimental setting. These combinations are denoted S, J or T , which are shortly called Bell parameters. In the Methods section in [49] it is explained that "*if the observed parameter violates the inequality, one can conclude that measured systems were not governed by any LHVM. It should be noted that this conclusion is always statistical, and typically takes a form of a hypothesis test, leading to a conclusion of the form: 'assuming nature is governed by local realism, the probability to produce the observed Bell inequality violation … is P(observed or stronger | local realism )≤p. This p-value is a key indicator of statistical significance in Bell Tests*."

Since p-values in several experiments are very small one concludes: "*Local realism, i.e., realism plus relativistic limits on causation , was debated by Einstein and Bohr using metaphysical arguments, and recently has been rejected by Bell tests*" . Such conclusion is imprecise and misleading. As Wiseman correctly explains in [44]:*"the usual philosophical meaning of "realism" is the belief that entities do exist independently of the mind, a worldview one might expect to be foundational for scientists."* It is also claimed that Bell tests allow rejecting *local causality*, where *Bell-local causality* is defined: Alice's output depends only on her input x and on $\lambda$ describing all possible common causes included in the intersection of the backward light cones of *a* and *b* and independent on inputs *x* and *y*.

It is true that tested probabilistic models have been motivated by various metaphysical assumptions. Nevertheless, Bell Tests allow only rejecting a statistical hypothesis saying that LHVM (15-17) provides a probabilistic coupling (20) consistent with experimental data. Therefore, the violation of Bell-CHSH inequalities does not allow for far reaching metaphysical speculations. We agree also with Hans de Raedt et al .[99]: *" all EPRB experiments which have been performed and may be performed in the future and which only focus on demonstrating a violation BI-CHSH merely provide evidence that not all contributions to the correlations can be reshuffled to form quadruples… These violations do not provide a clue about the nature of the physical processes that produce the data….*" Similar conclusions may be found in [21, 38, 62, 65, 72-76, 81, 100].

Bell Tests confirmed the existence of long range correlations between outcomes of experiments performed in space-like locations, If additional context dependent, variables, describing measuring instruments and procedures are correctly incorporated into the probabilistic model (11), then Bell-CHSH inequalities cannot be derived and "nonlocal " correlations can be explained without evoking quantum magic. We discuss such model in the next section.

### 9. Contextual hidden variable model and violation of statistical independence

We incorporate into the model (11) additional variables describing distant measuring contexts [84].

- $\lambda_1 \in \Lambda_1$ and $\lambda_2 \in \Lambda_2$ describe correlated physical systems and do not depend on measurement settings (x, y).
- $\mu_x \in M_x$ and $\mu_y \in M_y$ describe measurement procedures and instruments at the moment of measurement when settings (x, y) were chosen.
- Inputs/ labels (x, y) are randomly chosen in separate random experiment.
- Outputs are created locally: $a = A'_x(\lambda_1, \mu_x) = \pm 1$ and $b = B'_y(\lambda_2, \mu_y) = \pm 1$

The resulting contextual model (CHVM) is defined by three equations

$$E(A_x B_y) = \sum_{\lambda \in \Lambda_{xy}} A_x(\lambda_1, \mu_x) B_y(\lambda_2, \mu_y) P(\lambda_1, \lambda_2) P_{xy}(\mu_x, \mu_y) \qquad (21)$$

where $\Lambda_{xy} = \Lambda_1 \times \Lambda_2 \times M_x \times M_y$,

$$P(a, b, x, y) = \sum_{\lambda \in \Lambda_{xy}} P(a | \lambda_1, \mu_x) P(b | \lambda_2, \mu_y) P(\mu_x, \mu_y | x, y) P(x, y) P(\lambda_1, \lambda_2) \quad (22)$$

and

$$P(\mu_x, \mu_y | x, y) = P_{xy}(\mu_x, \mu_y) \neq P(\mu_x, \mu_y) . \quad (23)$$

In Bell Tests, P(x, y) = P(x) P(y), but in the contextual model (21) and in QM, it does not matter how labels (x, y) are chosen. In general spaces $\Lambda_{xy}$ for different settings (x, y) do not overlap and, as Larsson and Gill demonstrated [116], Bell-CHSH inequalities cannot be derived and $|S| \leq 4$.

The model (21-23) violates statistical independence and $P(x, y | \mu_x, \mu_y) \neq P(x, y)$:

$$P(\mu_x, \mu_y, x, y) = P_{xy}(\mu_x, \mu_y) P(x, y) = P(\mu_x, \mu_y) \to P(x, y | \mu_x, \mu_y) = 1 \quad (24)$$

The equation: $P(x, y | \mu_x, \mu_y) = 1$ "tells" only, that if a hidden event $\{\mu_x, \mu_y\}$ 'happened' then the settings (x,y) were used [14,82- 84, 111,112]. It has nothing to do with *conspiracy* and FoC.

Since inputs (x,y) were chosen using signals coming from distant stars [48], random number generators or random human choices made during online computer games [49], thus *fredom- of- choice -loophole* was successfully closed but it did not prove *statistical independence*. As we proposed in preceding papers a violation of *statistical independence* should be called *Bohr-contextuality* (which should not be cofounded with *CbD contextuality* [113-115] ) or simply *contextuality*.

CHVM violates Bell -locality and Bell-causality, but outputs are created in a locally causal way. Hidden variables describing physical systems and measuring contexts, in space-like separated laboratories, can be statistically correlated, but the violation of statistical independence and apparently nonlocal correlations may be explained without evoking spooky influences. It may be the effect of setting dependent post-selection of data or [37,82,84] it may due to the global space-time symmetries [83, 84,117].

The model (21) can be further simplified. For example $\mu_x$ can be a fixed set of variables describing experimental procedures labelled by *x*. If in a distant laboratory a setting labelled by *y* is used, then a measuring instrument and /or laser beam are rotated by an angle $\theta_{xy} = \theta_x - \theta_y$. Therefore, due to global rotational symmetry $\mu_y = f(\mu_x, \cos(\theta_{xy}))$ and :

$$E(A_x B_y) = \sum_{\lambda \in \Lambda_{xy}} A_x(\lambda_1, \mu_x) B_y(\lambda_2, f(\mu_x, \cos(\theta_{xy})) P(\lambda_1, \lambda_2) \quad (25)$$

The model (25) seems to have enough flexibility in order to explain the long range correlations in Bell Tests depending on $\theta_{xy} = \theta_x - \theta_y$ . The model (25) does not allow to derive any Bell-type inequalities.

## 10. Can Quantum-Mechanical Description of Physical Reality Be Considered Complete?

This question asked by Einstein, Podolsky and Rosen [32] and answered by Bohr [33] has been debated for 90 years and many incorrectly believe that the results of recent Bell Tests prove that if we reconcile QM with the general relativity we will obtain a complete description of physical reality. In fact we should be much more humble [118] because we even don't know whether QM is predictably complete.

QM gives probabilistic predictions for distributions of the results obtained in long runs of one experiment or in several repetitions of the same experiment on a single physical system. It is unclear how and in what sense a claim can be made that QM provides a complete description of individual physical systems. This is why Einstein [7, 8] never accepted that a statistical theory may provide a complete description of individual physical systems and believed that QM should be completed by some microscopic theory of sub-phenomena enabling to reproduce quantum probabilistic predictions.

According to Bohr, quantum probabilities describe completely quantum phenomena and experiments and no more detailed sub-quantum description is possible or necessary. In other words quantum probabilities are irreducible and QM is not an emergent theory. In statistical mechanics probabilities reflect the lack of knowledge about properties of physical systems. In SCI quantum probabilities reflect lack of knowledge, about interactions of physical systems with measuring instruments in well-defined experimental contexts. Bertrand paradox taught us that probabilities are not properties of individual physical systems but are only properties of random phenomena and experiments as a whole. In this sense they do not provide a complete description of individual physical systems.

Whether a more detailed description of quantum phenomena does exist, it is an open question and several hidden variable models were proposed and discussed. Bell Tests allowed rejecting several hidden variable models but neither proved completeness nor nonlocality of QM. Several years ago, we pointed out that the question about the completeness of QM cannot be answered by constructing ad hoc sub-quantum hidden variable models. It can only be answered by a different and a more detailed analysis of experimental data [11-13,78,119].

In quantum experiments, outcomes are registered by online computers as finite time series of data. It can be a laser beam which after passing by a PBS (polarization beam splitter) produces clicks on detectors coded ±1. It can be a physical system in a trap, a physical observable is measured an outcome is recorded and initial conditions in the trap are reset etc.

In all these experiments no single result is predictable. From long time-series of counts empirical frequency distributions are obtained and compared with probabilistic predictions of QT. In this way predictable *completeness* of QT is taken for granted and any fine structure of time- series, if it existed, would be averaged.

Let us consider two experiments repeated N times each. In the first experiment we obtain a time- series of the results:1,-1,1,-1,…1,-1… and in the second 1,-1,-1, 1,1,1,-1,1-1,-1,1,1,1,1,-1,… By increasing the value of N the relative frequency of getting 1 can approach 1/2 as close as we wish. However, it is not a complete description of these time-series. By searching for reproducible fine structures in experimental time-series we may investigate whether QM is emergent or not without constructing specific hidden variable models.

In any more detailed description of quantum phenomena, pure quantum ensembles become mixed statistical ensembles with respect to additional uncontrollable parameters describing physical systems and measuring instruments. There is a principal difference between a pure statistical ensemble and a mixed one. For a pure ensemble any sub-ensemble has the same properties. Sub-ensembles of a mixed statistical ensemble may differ from one to another if mixing is not perfect. These differences can be, in principle, detected by using so called purity tests [11, 12, 78, 119], which we introduced in a different context [120,121].

Let us consider a time -series of outcomes T(S, E, i) obtained in an i-th run of an experiment E performed on physical system(s) S. Since we do not control the distribution of hidden variables, the time-series T(S, E, i) may differ from run to run of the same experiment. Using the language of mathematical statistics the T(S, E, i) represents a random sample drawn from some statistical population. A pure ensemble is an ensemble characterized by such empirical distributions of various counting rates, which remain approximately unchanged for any rich sub ensembles drawn from this ensemble in a random way [12, 20, 78]. Therefore, we have to test the null hypothesis $H_0$:

*Samples T(S, E, i) for different values of i are drawn from the same statistical population*

Various statistical non-parametric compatibility tests can be used to test $H_0$.

Purity tests are not sufficient. To prove that QM is not *predictably complete* we have to study more in detail time-series of data, detect some temporal fine structure and find a stochastic model able to explain it. Several methods are used to study and to compare empirical time-series: frequency or harmonic analysis, period-grams, autocorrelation and partial autocorrelation functions etc. [122,123]. The aim of most of physical experiments is to compare empirical probability distributions with quantum probabilistic predictions. Therefore all fine structures in time- series of data if they existed are averaged out and not discovered.

Completeness of QM has been discussed for nearly hundred years but a detailed study of experimental time-series of existing experimental data is still waiting to be done. As we demonstrated recently with Hans de Raedt *sample inhomogeneity* invalidates significance tests [124], therefore if sample homogeneity is not tested carefully enough then *sample homogeneity loophole* is not closed and statistical inference cannot be trusted [124-126].

11. **Conclusions**

In this review article we explained why speculations about quantum nonlocality and quantum magic are rooted in incorrect interpretations of QM and/or in incorrect "mental pictures" and models trying to explain invisible details of quantum phenomena. In particular, it is not true that in Bell Tests entangled qubits behave as *"a pair of dice showing always perfectly correlated outcomes"*.

We advocate an abstract statistical contextual interpretation (SCI) of QM which is free of paradoxes. SCI rejects the existence of universal wave function. Quantum probabilities are objective properties of quantum phenomena. Whether these probabilities can be explained as emergent is an open question which cannot be settled by philosophical discussions and no-go theorems. It can be only elucidated by more detailed study of experimental time-series of data then it is usually done.

Bell Tests are subtle experiments being imperfect implementations of an ideal EPRB experiment. It is often claimed that violation of Bell-CHSH inequalities in these tests allowed rejecting with a great confidence *local realism* and *local causality*. We explained one more time, that such conclusions were misleading.

Bell-CHSH are trivial properties of Nx4 spreadsheets on which outcomes of measurements of 4 jointly distributed random variables e.g. $(A_x, B_y, A_{x'}, B_{y'})$ are displayed. In Bell Tests such experimental spreadsheets do not exist because we have 4 pairs of distant random experiments performed using 4 incompatible experimental settings (x, y). These experiments are described by empirical probability distributions of 4 pairs of random variables $(A_{xy}, B_{xy})$. Bell-CHSH inequalities cannot be derived and estimated pairwise expectations $E(A_{xy} B_{xy})$ are not constrained by these inequalities.

In order to explain statistical regularities in experimental data one can postulate probabilistic couplings for example: $E(A_{xy}B_{xy})=E(A_x B_y)$ etc. Quantum probabilistic model and Bell-causal hidden variable model can only be tested as plausible probabilistic couplings [84]. Quantum coupling (7) is constrained by quantum-CHSH inequalities: $|S| \leq 2\sqrt{2}$ [38, 74, 84, 127, 128]. Local hidden variable couplings (11,13, 15-17) are constrained by Bell-CHSH inequalities: $|S| \leq 2$.

It was incorrectly believed that if *freedom- of- choice* loophole was closed, then hidden variables could not statistically depend on randomly chosen binary inputs (settings' labels). It is not true because variables describing distant measuring instruments used in different settings can depend on the inputs and may be correlated due to the global rotational symmetry. Therefore, closing *freedom- of- choice* loophole does not close the *contextuality loophole*.

In contextual hidden variable models (21-23) and (25) distant outcomes are locally determined in function of setting independent hidden variables describing prepared qubits and setting dependent hidden variables describing distant measuring instruments and procedures. This model is only constrained by $|S| \leq 4$. Because of the global rotational symmetry pairwise expectation values of distant random variables (describing Alice's and Bob's outcomes) have to depend on an angle $\theta_{xy} = \theta_x - \theta_y$, where $(\theta_x, \theta_y)$ are, respective angles by which distant qubits are rotated before local read-outs.

Therefore, Bell Tests prove only that the probabilistic coupling LHVM is inconsistent with the experimental data. They allow rejecting Bell-locality and Bell-causality assumptions but they have not much to say about the completeness of QM or *local causality* in nature. As it was pointed out by several authors, *quantum nonlocality* is a misleading notion [60-65, 69-76, 81,83-85, 91-93,101, 98-101, 129,130] and extraordinary metaphysical speculations based on the results of Bell tests are unfounded.

We strongly believe that there exists an external world, which does not depend on whether it is observed or not and our mathematical models describe only imperfectly its different layers [118]. Quantum phenomena which investigate depend on detailed contexts of our experiments. The information we get is contextual and complementary but quantum probabilities are objective properties of quantum phenomena

The question about completeness of quantum mechanics can only be answered by a search of reproducible fine structures in time series of experimental data which were not predicted by QM. It would not only prove that QM may not provide the most complete description of the individual physical systems but it would also prove that QM is not *predictably complete* [11, 122, 123].

Let us finish this article with words of Einstein [8]:"*Is there really any physicist who believes that we shall never get any insight into these important changes in the single systems, in their structure and their causal connections…To believe this is logically possible without contradiction; but, it is so very contrary to my scientific instinct that I cannot forego the search for a more complete description*".

## 12. Conflict of Interest

*The authors declare that the research was conducted in the absence of any commercial or financial relationships that could be construed as a potential conflict of interest.*

### 13. Acknowledgments

.